\documentclass[apj]{emulateapj}

\usepackage[usenames,dvipsnames,svgnames,hyperref]{xcolor}
\usepackage{mathtools}
\usepackage{amsmath}
\usepackage{soul}
\usepackage{color}
\usepackage{natbib}
\usepackage{booktabs}
\defcitealias{lee13}{LY13}
\newcommand{\halomaker}{\textsc{HaloMaker}}
\newcommand{\gf}{\textsc{GF}}
\newcommand{\ramses}{\textsc{RAMSES}}
\newcommand{\ysamtm}{\textsc{ySAMtm}}
\newcommand{\yzics}{\textsc{YZiCS}}

\shorttitle{Wobbling galaxy spin axes in dense environments}
\shortauthors{Lee et al.}

\begin{document}
\title{Wobbling galaxy spin axes in dense environments}
\author{Jaehyun Lee\altaffilmark{1}, Suk Kim\altaffilmark{2}, Hyunjin Jeong\altaffilmark{2}, Rory Smith\altaffilmark{2}, Hoseung Choi\altaffilmark{3}, Ho Seong Hwang\altaffilmark{4}, Seok-Joo Joo\altaffilmark{2}, Hak-Sub Kim\altaffilmark{2}, Youngdae Lee\altaffilmark{2}, and Sukyoung K. Yi\altaffilmark{3}}
\altaffiltext{1}{Korea Institute for Advanced Study, 85, Hoegi-ro, Dongdaemun-gu, Seoul 02455, Republic of Korea, syncphy@gmail.com}
\altaffiltext{2}{Korea Astronomy and Space Science Institute, 776, Daedeokdae-ro, Yuseong-gu, Daejeon 34055, Republic of Korea}
\altaffiltext{3}{Department of Astronomy and Yonsei University Observatory, Yonsei University, Seoul 03722, Republic of Korea}
\altaffiltext{4}{Quantum Universe Center, Korea Institute for Advanced Study, 85, Hoegi-ro, Dongdaemun-gu, Seoul 02455, Republic of Korea}

\begin{abstract}

The orientation of galaxy spin vectors within the large scale structure has been considered an important test of our understanding of structure formation. We investigate the angular changes of galaxy spin vectors in clusters - denser environments than are normally focused upon, using hydrodynamic zoomed simulations of 17 clusters \yzics\ and a set of complementary controlled simulations. The magnitude by which galaxies change their spin vector is found to be a function of their rotational support with larger cumulative angular changes of spin vectors when they have initially lower $V_{\theta}/\sigma$. We find that both mergers and tidal perturbations can significantly swing spin vectors, with larger changes in spin vector for smaller pericentre distances. Strong tidal perturbations are also correlated with the changes in stellar mass and specific angular momentum of satellite galaxies. However, changes in spin vector can often result in a canceling out of previous changes. As a result, the integrated angular change is always much larger than the angular change measured at any instant. Also, overall the majority of satellite galaxies do not undergo mergers or sufficiently strong tidal perturbation after infall into clusters, and thus they end up suffering little change to their spin vectors. Taken as a whole, these results suggest that any signatures of spin alignment from the large scale structure will be preserved in the cluster environment for many gigayears.

\end{abstract}
\keywords{galaxies: evolution -- galaxies: elliptical and lenticular, cD -- galaxies: formation -- galaxies: stellar content }

\section{Introduction}
Galaxies acquire their primordial angular momentum via tidal torques which are induced by misalignment between the inertia tensor of their proto-halos and the local gravitational tidal tensor. Galaxies are thus expected to have spin vectors initially aligned with the principal axes of nearby large scale structures in their cosmological context~\citep{hoyle49,peebles69,doroshkevich70,white84,porciani02,casuso15,codis15b}. Assuming it is conserved, the initial angular momentum also determines the disk structures of galaxies in ideal infall models~\citep[e.g.][]{Mo98}. However, galaxies are expected to experience violent kinematic disturbances and mergers according to the hierarchical structure formation models in the $\Lambda$CDM paradigm~\citep[][and references therein]{white78,white79,ahmed81,miller83,villumsen83,duncan83,negroponte83,lake86,barnes92,hernquist93,quinn93,martel98,bournaud04,bournaud05,naab06,dimatteo09,jesseit09,shankar13,ji14,querejeta15,lagos18}. 

Numerical simulations show that the spin vectors are less aligned with nearby filaments when halos are more massive or involved in more mergers~\citep{navarro04,aragon-calvo07,brunino07,paz08,hahn10,codis12,libeskind12,aragon-calvo13,trowland13,zhang13,cen14,welker14,dubois14,codis15a,prieto15,zhang15,gonzalez17,liu17,wang18}. For example, ~\citet{codis12} propose that the weaker alignment is shown in larger dark matter halo masses because mergers are essential in building up massive halos while the winding of cosmic flows form small halos. Using a hydrodynamical zoom-in box simulation,~\citet{cen14} demonstrated that galaxy spin orientation is primarily determined by large-scale coherent torquing but it can be reoriented by local interactions. Meanwhile \citet{dubois14}, using the hydrodynamical cosmological-volume simulation Horizon-AGN, found that spin vectors are more closely aligned with nearby filaments for smaller, bluer, and more late type galaxies than for more massive, redder, and earlier type galaxies. 

Many observational attempts have been made to test the above-mentioned theoretical predictions~\citep{lee04,trujillo06,lee07,paz08,jones10,andrae11,varela12,tempel13a,tempel13b,zhang13,zhang15,hirv17,koo17}. Wide field surveys such as the Two Degree Field Galaxy Redshift Survey~\citep[2dFGRS,][]{colless01} and the Sloan Digital Sky Survey~\citep[SDSS,][]{york00} provide large and uniform data sets to test for a correlation between galaxy spin vectors and the large scale structure. The advent of integral field spectrographs enables us to directly derive the direction of galaxy spin vectors from spatially resolved kinematics of stellar and gas components in galaxies. For instance, \citet{emsellem07} found from the ATLAS$^{3D}$ IFU survey, for a volume-limited sample of 260 local early-type galaxies~\citep{cappellari11}, that early-type galaxies can be subdivided into slow and fast rotators. They claim that dissipation-less mergers may have twisted spin vectors and expelled angular momentum from slow rotators in dense environments~\citep[see also][]{lee18}.

As the densest regions in the Universe, galaxy clusters are found at the nodes of filamentary structures. The orientation of surrounding structures vary from cluster to cluster, indicating that the direction of spin vectors of cluster satellites can be distributed either randomly or anisotropically~\citep{peebles69,doroshkevich73,shandarin74,doroshkevich78,ozernoi78}. Moreover, galaxy clusters are the places where strong tidal fields can dynamically heat galaxies, and potentially induce morphological transformation ~\citep{moore96, gnedin97, moore98, gnedin03b,gnedin03a,park09,smith10,smith12b,smith16}. Using a set of cosmological hydrodynamical zoomed simulations for clusters, \citet[][]{choi17} found a significant decrease in the angular momentum of satellite galaxies in cluster environments. Some of this decrease is due to the actions of environmental effects, while galaxy mergers, alone, do not account for all the spin-down of galaxies in cluster environments. Therefore this could suggest that the orientation of galaxy spin vectors might also be affected by tidal interactions in clusters, in addition to mergers. 

Empirical studies have tried to connect the orientation of galaxy spins with the large scale structures in and around clusters. \citet{aryal04,aryal05} show that the spin vector orientation of galaxies in Abell clusters with respect to the local supercluster plane vary from cluster to cluster, with some being well aligned, while other clusters show the opposite or no preferred orientation~\citep[see also][]{hwang07}. On the other hand,~\citet{kim18} find a clear misalignment between two filamentary structures identified inside Virgo~\citep{west00,kraft11} and the spin vectors of bright early-types associated with the filaments, using the ATLAS$^{3D}$ IFU survey and the Extended Virgo Cluster Catalogue~\citep{kim14}. Therefore, it is still unclear whether the orientation of galaxy spin vectors is correlated with large scale structures after infall into clusters.

Such studies suffer from a fundamental limitation in that it is difficult to specify which structures the satellite galaxies came from. Furthermore, it is not straightforward to define the direction of large scale structures in dense environments which may be connected to multiple filaments simultaneously. So, instead we raise a more generalized question -- how much does the orientation of galaxy spin vectors change in dense environments, where strong tidal forces can perturb the kinematics of galaxies. In this study, we investigate this question using up-to-date cosmological hydrodynamical zoomed simulations, supplemented by a set of idealized simulations. 

This paper is organized as follows. In section 2, we briefly introduce our cosmological zoomed simulation and its post-processing for this investigation. In section 3, we look into the correlation between the angular changes of galaxy spin vectors and the multiple mechanisms that can arise, sometimes simultaneously, in cosmological zoomed simulations. Then, in Section 4, by using a set of idealized simulation, we delve into what parameters determine the impact of the cluster potential on the direction of galaxy spin vectors. We summarize our main results in Section 5.


\section{Cosmological zoomed simulation}
We use the \yzics\ simulation of 17 massive halos which are fully described in~\citet{choi17}. These simulations are conducted with the hydrodynamic code \ramses~\citep{teyssier02} using the cosmological parameters, derived from the 7-year Wilkinson Microwave Anisotropy Probe data~\citep{komatsu11}. This code has been updated to utilize the latest physical ingredients for the impact of baryons, such as AGN feedback and supernova feedback~\citep[see][]{dubois12}. Massive halos are selected from a cosmological volume of $200h^{-1}$Mpc cube, and zoom simulations are conducted on them. The final mass of the zoomed halos are in the range of $\log M_{200}/M_{\odot}\sim13.5-15.0$. The maximum refinement level of this simulation is 18, reaching up to $760h^{-1}$pc of cell resolution in the volume and the zoom region is chosen to be out to $3R_{200}$ of the halos at $z=0$.

\subsection{Generating galaxy catalogues and merger trees }

Galaxy catalogues are generated from the output snapshots of the \yzics\ simulation using a modified version of the halo finding code \halomaker~\citep{tweed09}, which is based on the AdaptaHOP technique~\citep{aubert04}. Gravitationally bound objects, composed of more than 200 stellar particles, are identified as galaxies. We construct galaxy merger trees from the galaxy catalogues by using the tree building code \ysamtm~\citep{jung14}. This code searches for the most likely descendants and progenitors of galaxies by comparing the identifications of stellar particles bound to galaxies between each snapshots. Galaxy merger trees can be composed of single or multiple branches. In this study, the branch linking the most massive galaxy at each time step is defined as the main branch of a galaxy merger tree, and the galaxies on the other branches are assumed to eventually merge with the main branch.

\subsection{Galaxy kinematics}

We measured the angular momentum of the stellar components of each galaxy. The orientation of the angular momentum is chosen to be the $z$-axis of cylindrical coordinate of each galaxy. In this study, the rotation velocity $V_{\theta}$ of a galaxy is the mean tangential velocity of stellar particles, within the effective radius $R_{\rm eff}$, the radius containing half the total stellar mass of a galaxy. The velocity dispersion of stellar components in a model galaxy $\sigma$ is defined to be $\sigma=\sqrt{(\sigma^{2}_{R}+\sigma^{2}_{\theta}+\sigma^{2}_{z})/3}$, where $\sigma_{R}$, $\sigma_{\theta}$, and $\sigma_{z}$ are the radial, tangential, and vertical velocity dispersion of stellar particles within $R_{\rm eff}$, respectively. To determine $\sigma_{\theta}$ at $r<R_{\rm eff}$, we first measured the tangential velocity dispersion of stellar particles found between $r$ and $r+\Delta r$, where $\Delta r=0.1R_{\rm eff}$, and averaged them to minimize the impact of the rotation curve on $\sigma_{\theta}$. We confirmed that $\Delta r=1/10R_{\rm eff}$ results in measurements that converge with those found using a smaller value of $\Delta r$. The ratio of the rotation velocity to the velocity dispersion $V_{\theta}/\sigma$ is used as a parameter to indicate the overall degree of rotational support in each model galaxy of the study.

\begin{figure}
\centering 
\includegraphics[width=0.5\textwidth]{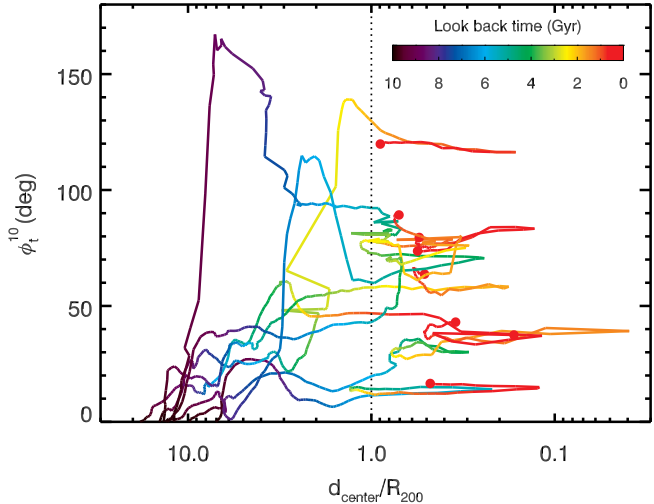}
\caption{Evolutionary tracks of 8 randomly chosen satellite galaxies in a net angular change-clustocentric distance space. The net angular change indicates the angle between spin vectors at 10 Gyr and $t$ in look-back time. The color code on the tracks denotes the look-back time. The filled red circles mark the final positions of the satellites on the space.}
\label{phi_d}
\end{figure}

\subsection{Angular changes of spin vectors}



\begin{figure}
\centering 
\includegraphics[width=0.5\textwidth]{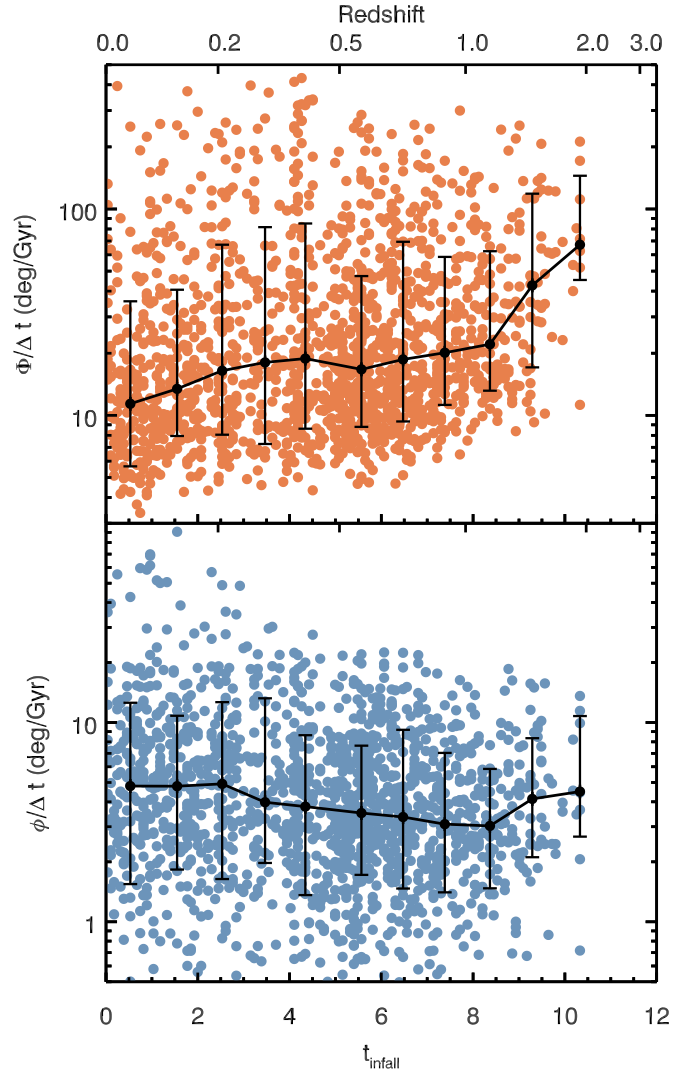}
\caption{Time averaged cumulative angular changes of spin vectors $\Phi/\Delta t$ (upper) and time averaged net angular changes of spin vectors $\phi/\Delta t$ (bottom), where $\Delta t=t_{\rm infall}-t_{z=0}$, as a function of infall epochs $t_{\rm infall}$. The black solid lines and the vertical bars present the median and $16^{\rm th}-84^{\rm th}$ percentile distributions of $\Phi/\Delta t$ and $\phi/\Delta t$ at given $t_{\rm infall}$. We note that the upper and bottom panels have different scales in the y-axis.}
\label{angle_dist}
\end{figure}

We take a cut in galaxy stellar mass at $10^{9.5} M_{\odot}$. This mass cut is adopted to ensure a reliable measurement of a galaxy's angular momentum without issues caused by the mass and spatial resolution of the simulation. For example, for this cut 97.5\% of galaxies at $M_{*}\sim10^{9.5} M_{\odot}$  have an effective radius that is four times larger than the spatial resolution of the simulation. The total number of galaxies above this mass cut within the zoom regions is 3038. Among them, 1629 satellite galaxies are found within  1$R_{200}$ of the 17 clusters at $z=0$. 

To mitigate numerical fluctuations that can sometimes arise on short timescales, the angular momentum of a galaxy at an epoch is averaged over $\pm0.25$ Gyr. We defined the net angular change of a galaxy's spin vector between two simulation time steps $i$ and $j$ as follows, 
\begin{eqnarray} 
\phi^{t_j}_{t_i}\equiv\frac{180}{\pi}\arccos \frac{\vec{L}(t_i)\cdot \vec{L}(t_j)}{|\vec{L}(t_i)||\vec{L}(t_j)|},
\label{eqn:tphi}
\end{eqnarray}
where $\vec{L}(t)$ is the angular momentum of a galaxy at time $t$, and $t_i$ and $t_j$ are the look-back time at $i^{\rm th}$ and $j^{\rm th}$ time steps. Figure~\ref{phi_d} displays the evolutionary tracks of 8 randomly chosen galaxies that finish up as cluster satellites, plotted in the plane of net angular change versus clustocentric distance (normalized by $R_{200}$). In this figure, the net angular change at any instant is measured with respect to its spin vector at a look-back time of 10 Gyr. The galaxy tracks show a range of behaviour before and after falling into the clusters.

Unlike the net angular change, the cumulative angular changes of the spin vector are a summation of the changes with time. This quantity enables us to parameterize the angular stability of spin vectors for a period of time $\Delta t=|t_j-t_i|$ by summing the angular changes between two serial snapshots from $i^{\rm th}$ to $j^{\rm th}$ time steps as follows,
\begin{eqnarray} 
\Phi^{t_j}_{t_i}\equiv\sum_{n=i}^{j-1} \phi^{t_{n+1}}_{t_n}.
\label{eqn:tphi}
\end{eqnarray}
We note that, in some circumstances, the net and cumulative changes in spin vector may deviate significantly with time. For example, consider the case where, in a first step, a galaxy changes its spin vector and then, in a second step, suffers a reversal of that change. The cumulative change will sum up both changes where as the net may return to the value it had before the first step. As such, if we are looking to see if galaxies maintain their pre-infall spin vectors in the cluster environment, the net change may be a better indicator for comparison with the observations, as it shows the current alignment with respect to an earlier one. However, the cumulative change may be more useful for quantitatively understanding the impact of dense environments on spin vectors within the simulation.

   \begin{figure*}
\centering 
\includegraphics[width=0.85\textwidth]{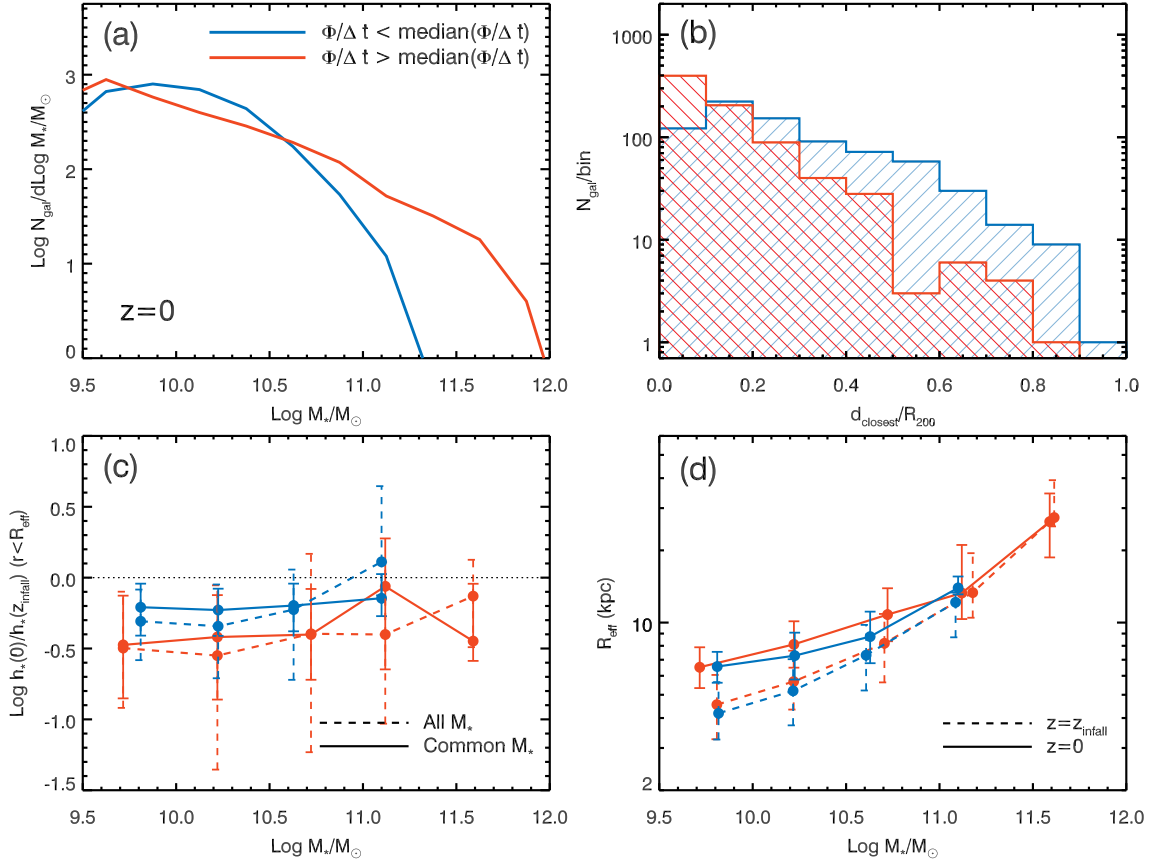}
\caption{Comparison of galaxy properties between the two groups of satellite galaxies classified by their time averaged cumulative angular changes of spin vectors $\Phi/\Delta t$. The color code indicates the same to that in Figure~\ref{v_sigma}. (a) Stellar mass functions of the two groups at $z=0$. (b) The histogram of galaxies found to be satellites in clusters at $z=0$ as a function of the closest clustocentric distances over $R_{200}$. (c) The ratios of the specific angular momentum of stellar mass within $R_{\rm eff}$ at $z=0$ to those at $z_{\rm infall}$ as a function of final stellar mass. The dashed lines are for all stellar particles and the solid lines are for the stellar particles commonly found within $R_{\rm eff}$ at $z=0$ and $z_{\rm infall}$. The vertical bars mark $16^{\rm th}-84^{\rm th}$ percentile distributions of the specific angular momentum ratios. (d) Stellar mass-size relation of the two groups at $z=0$ (solid) and $z_{\rm infall}$ (dashed). }
\label{gal_prop}
\end{figure*}

In this study, we mainly focus on the spin vector changes of satellite galaxies after falling into clusters. Therefore, $\phi^{t_{\rm infall}}_{t_{z=0}}$ and $\Phi^{t_{\rm infall}}_{t_{z=0}}$, where $t_{z=0}$ is the epoch at $z=0$ and $t_{\rm infall}$ is the epoch when the satellites are first crossing the virial radii of their final host halos, are reduced to $\phi$ and $\Phi$ for simplicity.

Figure~\ref{angle_dist} shows the time averaged cumulative angular changes and time averaged net angular changes of satellite spin vectors $\Phi/\Delta t$ and $\phi/\Delta t$, where $\Delta t=t_{\rm infall}-t_{z=0}$, plotted as a function of $t_{\rm infall}$ in look-back time. The median $\Phi/\Delta t$ (top panel, cumulative change) tends to steadily increase with time since infall (the median of the total sample $\sim17$ deg/Gyr). Meanwhile $\phi/\Delta t$ (bottom panel, net change) is more constant with time since infall.  Also, the net values are typically lower than the cumulative values (the median of the total net sample is $\sim4$ deg/Gyr compared to $\sim17$ deg/Gyr for the total cumulative sample). In most galaxies, spin vectors swing in a varying direction with time, and thus end up canceling out three quarters of their angular changes over time. We try to estimate the contribution of non-physical processes to $\Phi/\Delta t$ by using a control sample. This control sample is made up of isolated galaxies that suffer only weak external tidal perturbations ($\log p(t) < -5$; see Section 3.4) over the relatively tranquil period of their evolution since they formed half their final stellar mass. We also consider only stars formed prior to that period, so as to minimise the potential impact of newly formed stars in changing the spin vector. Finally, the sample is selected to have no recent galaxy mergers occuring within 1~Gyr of the current snapshot. By measuring angular change rates of spin orientation for this sample, we estimate that non-physical processes contribute only $\sim$ 5 deg/Gyr to $\Phi/\Delta t$ in our main sample. These differences clearly highlight that {\it the angular changes of spin vectors after cluster infall are dominated by oscillating or wobbling of spin vectors, often canceling out previous changes.} Among the cluster satellites, $\sim72\%$ of them have a net change, $\phi$, less than 30 deg after infall into the cluster environments. This already demonstrates that, in most cases, pre-infall spin vectors may be well preserved in the cluster environment on long timescales. 

In the rest of our analysis, we use only galaxies with $t_{\rm infall}>0.5$ Gyr to reduce the bias that could be introduced by small $t_{\rm infall}$ (i.e. overestimation of the time averaged properties). Among 1629 satellites, 1547 galaxies have $t_{\rm infall}$ older than 0.5Gyr.

\section{Spin vector changes in cosmological zoomed simulation}

\subsection{Physical properties of satellite galaxies depending on angular changes of spin vectors}

In this section, we correlate the angular change in galaxy spin vectors with other galaxy properties to try to find out which mechanisms induce the orientation change. We divide the satellite galaxies into two groups (i.e. those with a time averaged cumulative angular change in spin vector $\Phi/\Delta t$ below and above the median, 17deg/Gyr), and compare the physical properties between the two groups. Figure~\ref{gal_prop} displays the stellar mass functions, the histogram as a function of the closest clustocentric distances normalized by $R_{200}$, the magnitude changes of the specific angular momentum, and the mass-size relation of the two groups of satellites. 

Panel (a) shows a clear difference in stellar mass function between the two groups. The galaxies with lower $\Phi/\Delta t$ are dominated by those with $\log M_{*}/M_{\odot}<11$, and none of them is found above $\log M_{*}/M_{\odot}\sim11.3$. On the other hand, galaxies with higher $\Phi/\Delta t$ have stellar masses over the entire mass range. This implies larger satellite galaxies are more likely to have relatively unstable spin vectors. Indeed, galaxy mass is already known to correlate with many other galaxy properties including galaxy morphologies, colors, star formation history and spin parameters \citep[e.g.][]{cervantes-sodi08,calvi12,alpaslan15,moffett16,poudel16}. Panel (b) shows the distribution of the closest clustocentric distances normalized by $R_{200}$ (i.e. $d_{\rm closest}/R_{200}$) for the two groups. We note that a galaxy which merges with the cluster might end up with a very small minimum clustocentric distance. However, as we only consider galaxies that survive until z=0, these objects are automatically exlcuded from our analysis. Overall, the galaxies with higher $\Phi/\Delta t$ approach the cluster centers more closely than those with the lower $\Phi/\Delta t$. This can be naturally explained as the result of galaxies with smaller clustercentric distances suffering stronger perturbations, resulting in larger spin vector changes. This suggests that the orbital parameters of satellite galaxies could be one of important parameters influencing spin vector stability. This will be discussed further in \S4.2. using a set of controlled simulations.

\begin{figure}
\centering 
\includegraphics[width=0.45\textwidth]{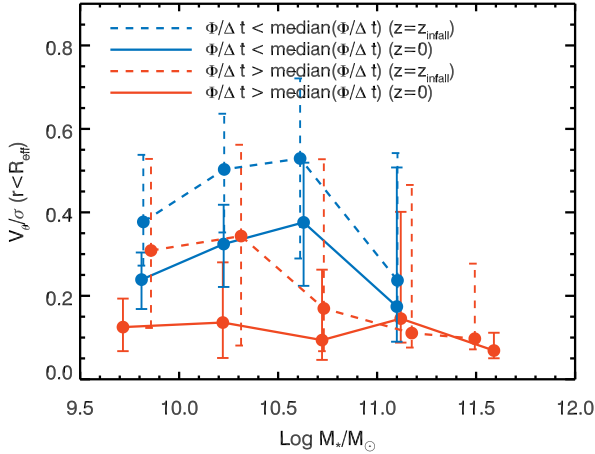}
\caption{$V_{\theta}/\sigma~(r<R_{\rm eff})$ of satellite galaxies as a function of galaxy stellar mass. The red and blue denote the groups of satellites having $\Phi/\Delta t$ above and below the median of $\Phi/\Delta t$, which is $\sim8.04$ deg/Gyr. The solid lines present the medians of $V_{\theta}/\sigma$ of the galaxies binned by their stellar mass at $z=0$ and the dashed lines are for their main progenitors at $z=z_{\rm infall}$. The vertical bars show $16^{\rm th}-84^{\rm th}$ percentile distribution of $V_{\theta}/\sigma$ at given stellar mass.}
\label{v_sigma}
\end{figure}

Panel (c) shows the ratios of the specific angular momentum of stellar components within $R_{\rm eff}$ at $z=0$ to that at $z=z_{\rm infall}$ as a function of total stellar mass. The dashed lines are for all stellar components within $R_{\rm eff}$ and the solid lines are for those star particles found within $R_{\rm eff}$ at both epochs. By comparing the two samples, we can attempt to differentiate what angular momentum change occurs due to tidal stripping removing stars and thus carrying away angular momentum, and what angular momentum change occurs due to the stirring up of the existing stars. The stars that exist at both epochs can be seen to lose their specific angular momentum in the cluster environment, which demonstrates that tidal stirring must play a role. This decrease is more significant for the galaxies with higher $\Phi/\Delta t$. The same is true for the total stellar component sample except that the specific angular momentum loss is generally slightly larger, except at the most massive end. This suggests that tidal stripping may play an additional role on top of tidal stirring, or that the stellar components that are accreted into $r<R_{\rm eff}$ after infall are those undergoing angular momentum loss more severely than the common components for the same period of time.

\begin{figure*}
\centering 
\includegraphics[width=1\textwidth]{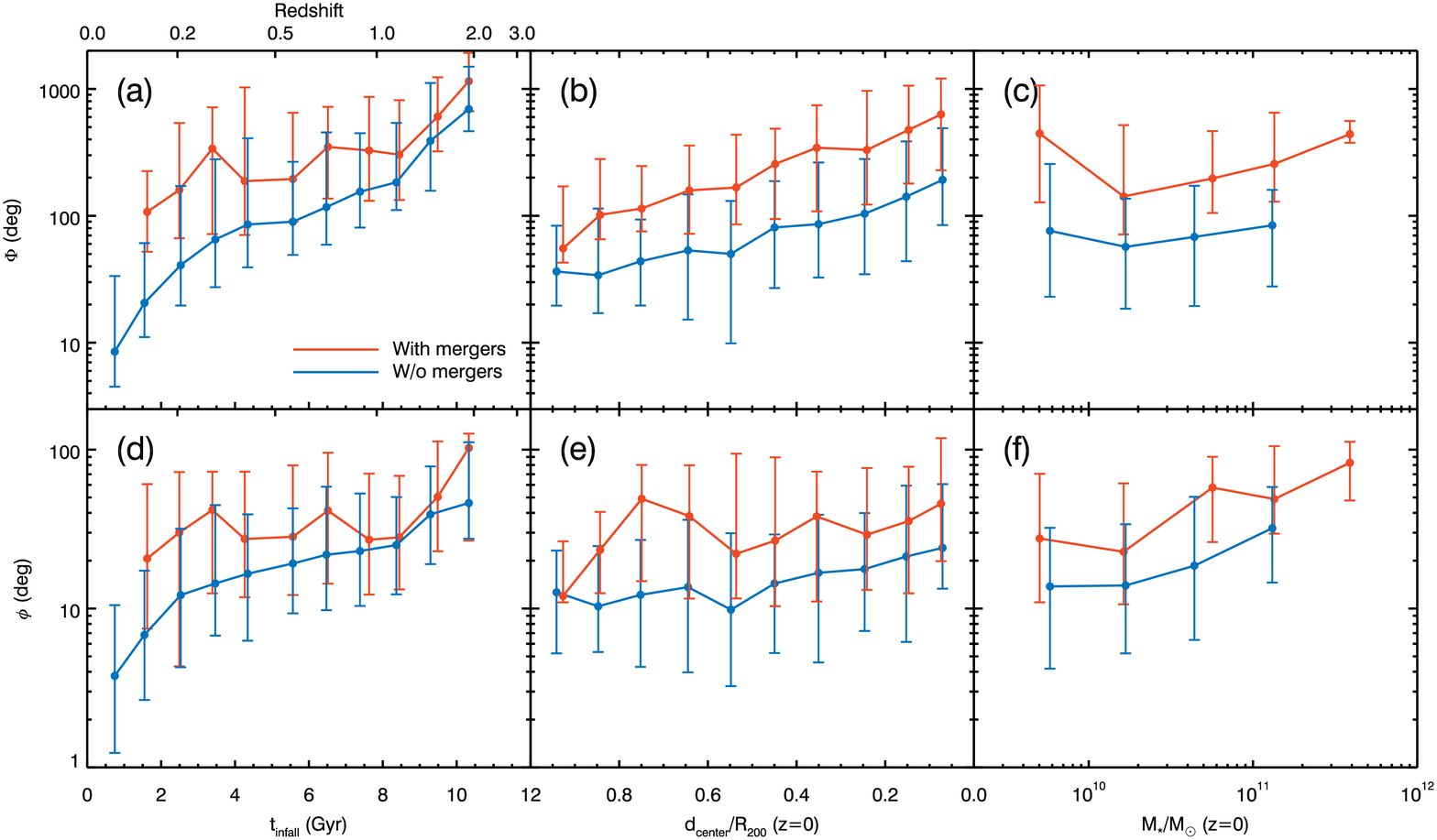}
\caption{Cumulative angular changes (upper) and net angular changes (bottom) of galaxy spin vectors at given infall time (left), clustocentric distances normalized by $R_{200}$ at $z=0$ (middle), and stellar mass at $z=0$ (right). The red indicates galaxies experiencing mergers of $\mu>0.1$ after infall and the blue is for those that do not undergo the mergers after infall. The vertical bars mark $16^{\rm th}-84^{\rm th}$ percentile distribution. We note that the clustocentric distance decreases from left to right in the middle panels. We note that the upper and bottom panels have different scales in y-axis.}
\label{merger_infall}
\end{figure*}

On the other hand, the mass-size relation in Panel (d) show no significant differences between the two groups at the two different epochs. This indicates no correlation between spin vector changes and galaxy size. Meanwhile, the gaps between the two epochs become smaller with increasing stellar mass. This is mainly because more massive galaxies found inside clusters at $z=0$ are those which fell into clusters more recently. More massive galaxies essentially have shorter merging time scales due to their larger initial halo mass~\citep[e.g.][]{Binney87,boylan-kolchin08}, and thus those that fall into clusters earlier tend to merge more quickly with the cluster centrals.

\subsection{Stability of spin vectors and $V_{\theta}/\sigma$}
Intuitively, we might expect that stronger torques are required to swing a system with larger angular momentum. If so, the change in spin vector direction would be expected to anti-correlate the amount of rotational support in a galaxy's stars. We parameterize the degree of galaxy rotation with $V_{\theta}/\sigma$ (i.e. ratio of rotational velocity to velocity dispersion) of the stellar components at $r<R_{\rm eff}$ for our model satellite galaxies. Figure~\ref{v_sigma} displays $V_{\theta}/\sigma$ as a function of stellar mass at $z=z_{\rm infall}$ (dashed) and $z=0$ (solid). As in Figure~\ref{gal_prop}, a comparison is made between the above- and below-the-median-$\Phi/\Delta t$ groups in this figure. 

Most galaxies, regardless of their stellar mass, show a decrease in $V_{\theta}/\sigma$ after infall, as also demonstrated in~\citet{gnedin03a}, ~\citet{choi17} and~\citet{choi18}. At a given mass, galaxies with higher $\Phi/\Delta t$ have $V_{\theta}/\sigma$ smaller than those with lower $\Phi/\Delta t$ at both $z=0$ and $z_{\rm infall}$. This suggests that the spin vectors are more easily reoriented in more pressure supported systems. Thus, initial $V_{\theta}/\sigma$ could be another important parameter correlated with the orientation stability of galaxy spin vectors. Meanwhile, at fixed $V_{\theta}/\sigma$, more massive galaxies have larger angular momentum, resulting in less changes in the direction of their spin vectors for galaxies with $\log M_{*}/M_{\odot} <11$. Galaxy clusters generally prevent the satellite galaxies from gas accretion while also inducing gas stripping. Therefore, satellite galaxies find it challenging to rejuvenate their rotational support with additional star formation, and instead experience a decrease in $V_{\theta}/\sigma$ after infall. Massive galaxies ($\log M_{*}/M_{\odot}>11$) in both groups commonly have $V_{\theta}/\sigma$ lower than 0.2 at infall epoch, because they build up their mass mainly via mergers, ending up having pressure supported kinematics before infall~\citep{oser10,lee13,dubois13,hirschmann15,rodriguez-gomez16,dubois16,lee17}. Given that specific angular momentum (e.g. panel (c) of Figure~\ref{gal_prop}) and $V_{\theta}/\sigma$ (e.g. Figure~\ref{v_sigma}) are both measures of the amount of rotation in a galaxy, these results combined demonstrate that spin orientation changes are closely correlated with spin amplitude changes.

The horizontal difference in position of the symbols on the dashed and solid lines (for lines of matching color) indicates how galaxies have changed their stellar mass since infall. The massive galaxies ($\log M_{*}/M_{\odot}>11.5$) have increased their mass since $z=z_{\rm infall}$ while the less massive ones tend to lose their stellar mass after infall. This is because most galaxies in the massive end were actually centrals of galaxy groups before entering the cluster and, as such, can continue to grow through cannabalising their subgroups while in the cluster. On the other hand, the less massive galaxies tend to lose mass due to tidal stripping being more effective than mass growth via star formation or mergers.

\subsection{Impact of mergers}

Galaxy mergers can violently disturb the kinematics of galaxies, in particular for major mergers. Thus, we might expect that both the cumulative and net angular change in spin vectors to be boosted by mergers. In this study, we consider only the cases where the mass ratio $\mu>1/10$, which is expected to be significant enough to have some impact on the galaxy dynamics. Figure~\ref{merger_infall} demonstrates the impact of mergers on $\Phi$ and $\phi$ as a function of $t_{\rm infall}$ (left panels), the fractions of clustocentric distances at $z=0$ (middle panels) and stellar mass at $z=0$ (right panels). As expected, mergers lead to an increase in $\Phi$ and $\phi$ by roughly a factor of two compared to the cases without mergers. This is comparable to the role of mergers in the magnitude of the galaxy spin changes from \citet{choi17}, where roughly half of galaxy spin change is due to mergers.

Panel (b) shows that $\Phi$ increases with decreasing $d_{\rm center}/R_{200}$ at $z=0$ despite non-negligible scatter in the trend. This is because satellite galaxies that fell in earlier tend to be found closer to the cluster centre at $z=0$~\citep{rhee17}, and thus the gradient is primarily driven by the stronger trend with infall time seen in the left hand panel. However, this is not so clearly the case with $\phi$; a much weaker correlation is buried in large scatter. Indeed, $\Phi$ is twice as strongly correlated with $d_{\rm center}/R_{200}$ than with $\phi$ in both subsets (i.e. with or without mergers). For example, the Pearson's correlation coefficient is -0.46/-0.34 for galaxies with/without mergers in the $\Phi-d_{\rm center}/R_{200}$ relation, and -0.10/-0.17 for galaxies with/without mergers in the $\phi-d_{\rm center}/R_{200}$ relation. The differences between $\Phi$ and $\phi$ thus become larger closer to the cluster centre. Once again, like in Figure 2, this indicates that the angular changes of spin vectors are dominated by precession or wobbling, but often those changes later cancel out. The majority of galaxies show net angular changes $\phi$ less than 30 deg even at $d_{\rm center}/R_{200}\sim0.1$ without mergers.

The difference between panels (b) and (e) once again shows that the orientation of spin vectors established before falling into clusters can be maintained for a long period after cluster infall. The cluster environment does not necessarily wipe out the pre-infall orientations, and as a result we do not see a strong clustocentric radial trend, and we would expect an even weaker radial trend in observed clusters where distances from the cluster centre are only projected distances. This result also implies that the cluster-to-cluster variation of the orientation of galaxy spin vectors found by~\citet{aryal04,aryal05} probably originates from varying large scale structures around clusters.

The right panels show the relation between the final stellar mass and the angular changes of spin vectors. A notable feature is the different dependence of $\Phi$ on stellar mass compared to $\phi$. $\Phi$ shows a weak correlation with final stellar mass, but $\phi$ shows a much stronger correlation. Most of the with-merger galaxies have $\Phi$ larger than 100 deg while their $\phi$ is $\sim30$ deg at $\log M_*/M_{\odot}\sim 10$ and spin vectors end up being perpendicular to their initial orientation in the massive end ($\log M_*/M_{\odot}>11$). This indicates that spin vector changes are dominated by precession or wobbling within a limited angular range in smaller galaxies, meanwhile the changes are more drastic in massive galaxies. This is likely related to the fact that massive galaxies tend to have an initially lower $V_{\theta}/\sigma$ seen in Figure~\ref{v_sigma}.

\begin{figure}
\centering 
\includegraphics[width=0.5\textwidth]{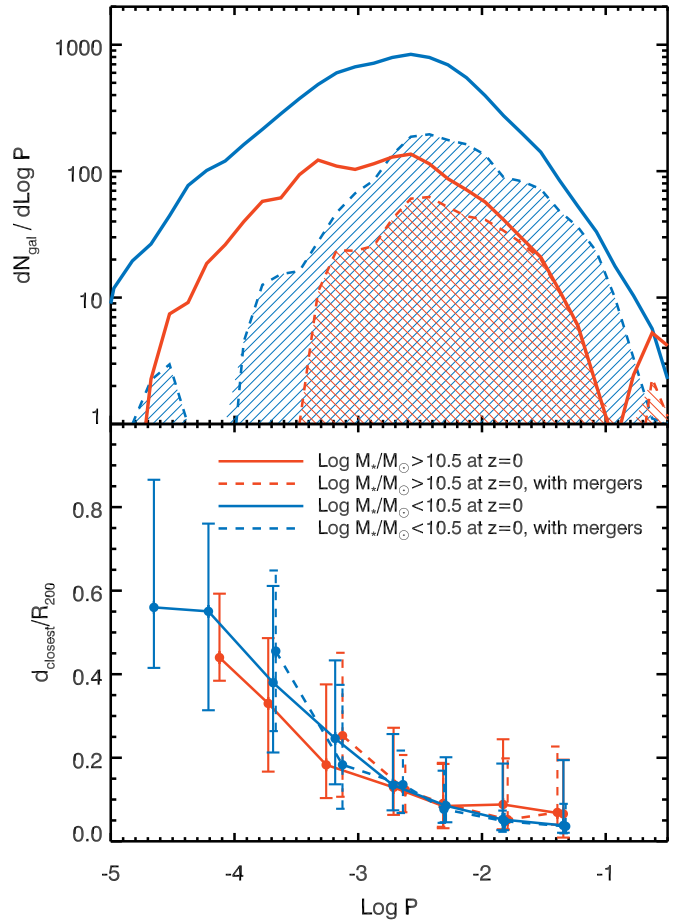}
\caption{Number density of galaxies as a function of time averaged perturbation index (upper) and $d_{\rm closest}/R_{200}$-to-$\log P$ relation (bottom). The red and blue denote the groups of galaxies that have stellar mass of $\log M_*/M_{\odot}>10.5$ and $\log M_*/M_{\odot}<10.5$ at $z=0$. In the upper panel, the hatched regions below the dashed lines mark the number density of galaxies that are involved in mergers with $\mu>1/10$ after infall. In the bottom panel, the solid and dashed lines indicate all galaxies and those involved in mergers after infall in a mass group. The vertical bars show $16^{\rm th}-84^{\rm th}$ distributions of $d_{\rm closest}-R_{200}$ at given $\log P$ }
\label{perturbation_hist}
\end{figure}

\subsection{Impact of perturbation}

We have examined the correlation between the angular change in spin vectors with individual galaxy properties and their merger histories.  As Figure~\ref{merger_infall} demonstrates, the galaxy spin vectors are notably swung even without any significant mergers. Another likely physical process disturbing spin vectors is tidal perturbation. In this section, we investigate the impact of tidal perturbation on the angular changes of the spin vectors of satellite galaxies.

Panel (b) of Figure~\ref{gal_prop} implies a correlation between environments and the orientation changes of galaxy spin vectors. Satellite galaxies experience stronger tidal force as approaching further inside of cluster potential. Therefore, the ratio of the closest clustocentric distances normalized by $R_{200}$ of satellite galaxies $d_{\rm closest}/R_{200}$ can be used as a proxy indicating the degree of tidal perturbation induced by the gradient of cluster potential.  

Galaxies are located at the local minima of the gravitational potential fields, and thus one can estimate the degree of tidal perturbation caused by fluctuating potential fields from the distribution of neighboring galaxies.~\citet{byrd92} propose a perturbation parameter which enables one to make a relative comparison of the strength of tidal perturbation exerted by a distant object onto a galaxy. The perturbation parameter is formulated as $(M_P/M_g)(R/d)^3$, where $M_P$ is the mass of a perturber, $M_g$ is the mass of a perturbed object, $R$ is the size of the perturbed object, and $d$ is a distance from the perturber to the perturbed object. This formula can be re-written as $(R^2/GM_g)(GM_P R/d^3)$, in which the first term is the inverse of the magnitude of the gravitational acceleration due to $M_g$ at $R$ and the second term is the magnitude of the tidal acceleration induced by $M_P$ at $R$ of the perturbed object. Therefore, the perturbation parameter of~\citet{byrd92} can be interpreted as the ratio of the tidal acceleration to the gravitational acceleration of the perturbed object exerted on a unit mass at $R$. In this sense, we parameterize the mean impact of the tidal force upon a galaxy $i$ due to the density fields from the galaxy distribution for a period of time $\Delta t=t_1-t_0$ as follows,
\small
\begin{equation}
\label{eqn:perturbation}
\begin{split}
\log P_{i}(t_{0},t_{1})\equiv\log \frac{1}{\Delta t}\int_{t_{0}}^{t_{1}}p_{i}(t)dt \\
 =\log \frac{1}{\Delta t}\int_{t_{0}}^{t_{1}} \frac{R_i^2(t)}{M_i(t)} \bigg| \sum_{j=1}^{j\neq i} \frac{M_j(t) R_i(t)}{d_{ij}^3(t)}\vec{u}_{ij} (t) \bigg|  dt,
\end{split}
\end{equation}
\normalsize
where $M_j$ is the stellar mass of the $j$th galaxy among $n$ galaxies, which are assumed to act as perturbers in a simulation volume, $M_i$ is the stellar mass of the perturbed galaxy, $R_i$ is the radius of the perturbed galaxy, $d_{ij}$ is a distance between the $i^{\rm th}$ and $j^{\rm th}$ galaxies, and $\vec{u}_{ij}$ is the unit vector in the direction between the $i^{\rm th}$ and $j^{\rm th}$ galaxies. According to Eq.~\ref{eqn:perturbation}, a satellite galaxy that survives until z=0 in a cluster experiences a mean perturbation of $\log P(t_{\rm infall},t_{z=0})$ after infall. For simplicity, $\log P(t_{\rm infall},t_{z=0})$ is reduced to $\log P$ hereafter. We note that the cluster potential is excluded from the summation term. In this study, the effective radius of a galaxy is adopted as the radius of the perturbed galaxy (i.e. $R_i$), and, accordingly, half of the total stellar mass is adopted as $M_i$. This formula gives the mean ratio of the net tidal acceleration to the gravitational acceleration of a perturbed galaxy after infall. This parameter is also used to indicate the mean tidal perturbation of galaxies caused by a density field after infall, along with $d_{\rm closest}/R_{200}$.

Figure~\ref{perturbation_hist} displays the distribution of perturbation index $\log P$ for the sample of galaxies (top panel) and the correlation between $d_{\rm closest}$ and $\log P$ (bottom panel). We divide the galaxies into four groups based on their final stellar mass (i.e. $\log M_*/M_{\odot}>10.5$: red and $\log M_*/M_{\odot}<10.5$: blue) and on whether they suffer mergers (i.e. with mergers: dashed, without mergers: solid). The hatched regions below the dashed lines in the top panel lean toward higher $\log P$ compared with overall number density, showing that the galaxies that have experienced mergers tend to have higher $\log P$. This could partly be caused by fact that merger counterparts boost the index during the merging phase, as they orbit closely to their central galaxies before merging. The two groups with different stellar masses show similar distributions of $\log P$, indicating negligible dependence of $\log P$ on the final stellar mass. The fraction of galaxies experiencing mergers is higher when $\log M_*/M_{\odot}>10.5$, consistent with the results in previous studies~\citep[e.g.][]{lee13}. The bottom panel shows a strong correlation between $d_{\rm closest}/R_{200}$ and $\log P$. {\it This indicates that the perturbing galaxies are more concentrated near the central regions of clusters. where number density of galaxies is higher}.

\begin{figure}
\centering 
\includegraphics[width=0.5\textwidth]{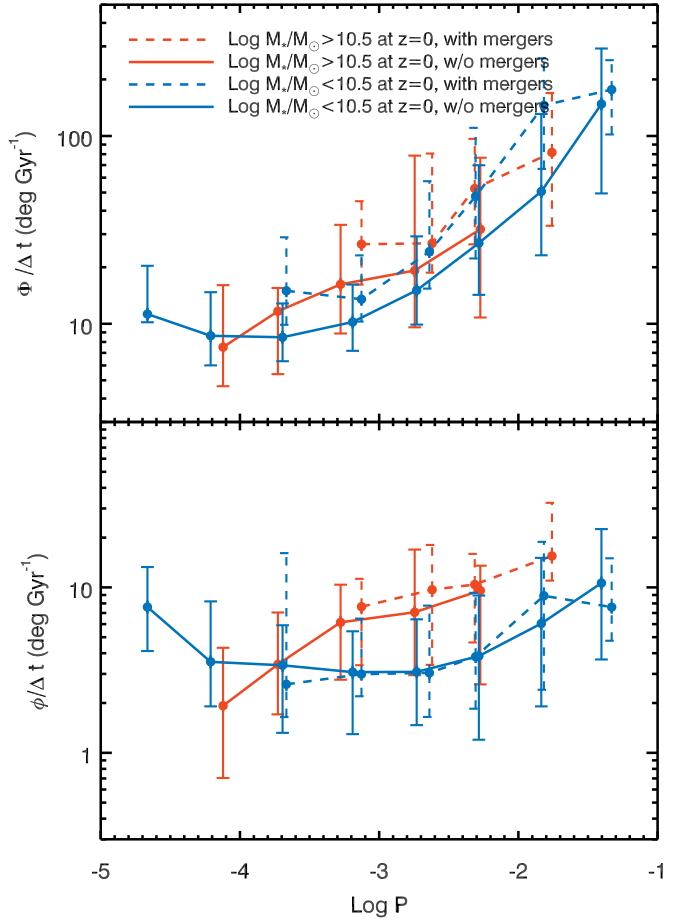}
\caption{Time averaged cumulative angular changes of spin vectors $\Phi/\Delta t$, where $\Delta t=t_{\rm infall}-t_{z=0}$, and net angular changes of spin vectors $\phi/\Delta t$ as a function of mean perturbation index $\log P$. The color code denotes final stellar mass and the different line styles indicate the cases with/without mergers (dashed/solid) after infall with $\mu>1/10$. The vertical bars and filled circles present $16^{\rm th}-50^{\rm th}-84^{\rm th}$ percentile distributions. We note that the upper and bottom panels have different scales in y-axis.}
\label{perturbation1}
\end{figure}

\begin{figure*}
\centering 
\includegraphics[width=1\textwidth]{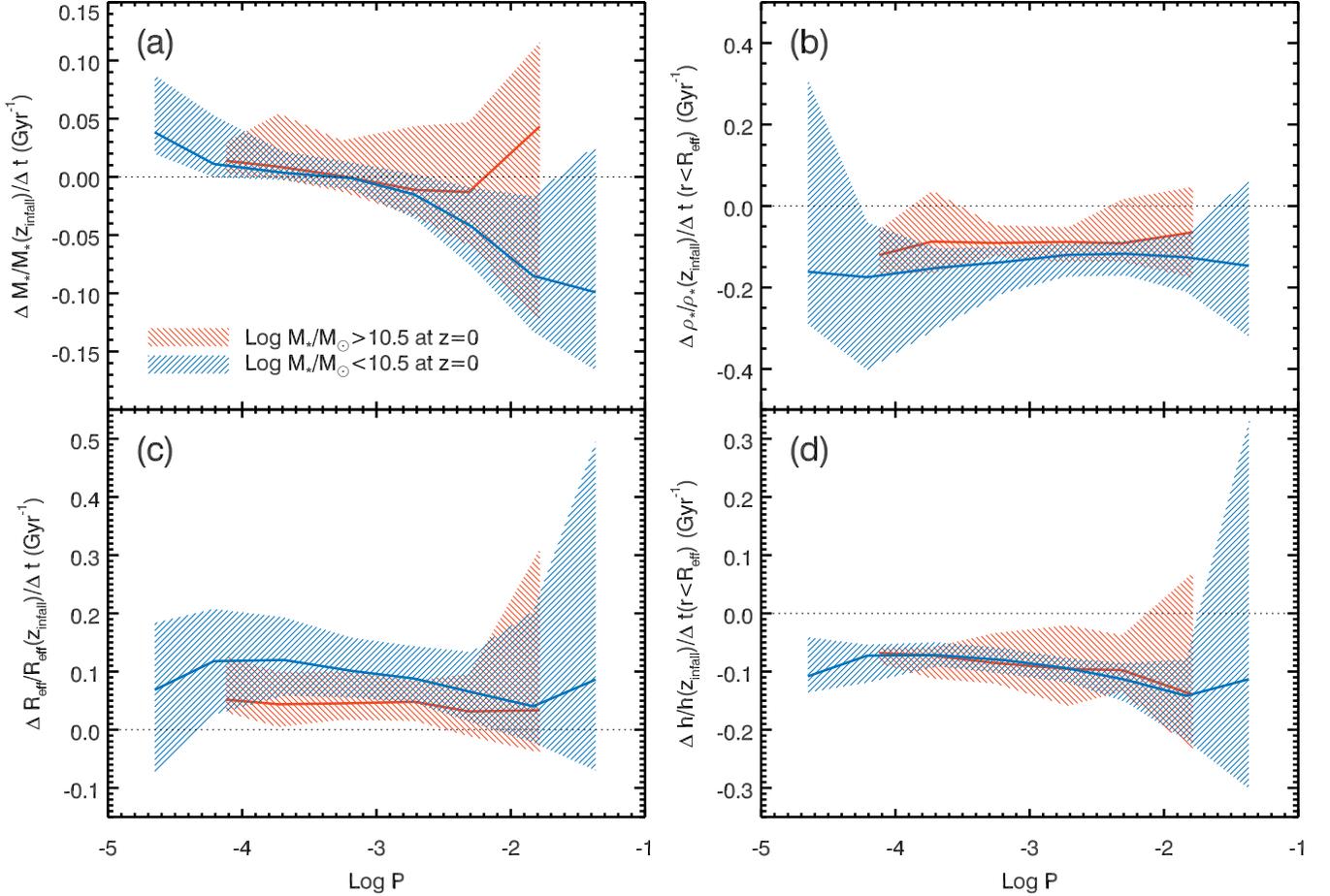}
\caption{Specific change rates of (a) galaxy stellar mass, (b) stellar mass density within $R_{\rm eff}$, (c) $R_{\rm eff}$, and (d) the magnitude of specific angular momentum $|h|$ within $R_{\rm eff}$ after infall into clusters as a function of the time average perturbation index. The red and blue denote galaxies with $\log M_{*}/M_{\odot}>10.5$ and $\log M_{*}/M_{\odot}<10.5$ of final mass and the dashed and solid lines mark those with/without mergers of $\mu>1/10$ after infall. The vertical bars present $16^{\rm th}-84^{\rm th}$ percentile distribution of each case at given $\log P$. The hatched region shows $16^{\rm th}-84^{\rm th}$ percentile distribution of all galaxies in each mass range.}
\label{perturbation2}
\end{figure*}

\subsubsection{Dependence of spin vector changes on $\log P$}

We investigate the impact of perturbation on the orientation changes of galaxy spin vectors in this section. Figure~\ref{perturbation1} shows the time averaged cumulative angular changes of galaxy spin vectors $\Phi/\Delta t$, where $\Delta t=t_{\rm infall}-t_{z=0}$, and time averaged net angular changes of spin vectors $\phi/\Delta t$ as a function of tidal perturbation index $\log P$. The color code and line styles are the same as those in the bottom panel of Figure~\ref{perturbation_hist}. 

In the top panel, $\Phi/\Delta t$ shows a strong correlation with $\log P$ for all cases, while in the bottom panel $\phi/\Delta t$ shows only a weak dependence. Quantitatively, the $\Phi/\Delta t-\log P$ relation has a Pearson's correlation coefficient larger than that of $\phi/\Delta t-\log P$ relation: 0.68 for $\Phi/\Delta t$ compared to 0.20 for $\phi/\Delta t$. This is because galaxies experience a \st{net} tidal force that varies in direction as they orbit within a cluster. The tidal perturbation swings spin vectors in varying directions, leading to the strong correlation between $\Phi/\Delta t$ and $\log P$. However, its net effect is largely cancelled out, resulting in the difference between $\phi/\Delta t$ and $\Phi/\Delta t$. This means that the angular change of spin vectors are dominated by wobbling within a limited angular range. 

Figure~\ref{merger_infall} showed the significant impact that mergers have on the angular changes of spin vectors after infall. However, unlike in Figure~\ref{merger_infall}, Figure~\ref{perturbation1} shows that the angular changes are not strongly affected by the mergers at a given $\log P$. Time averaged cumulative angular changes of spin vectors $\Phi/\Delta t$ of both the groups strongly depends on $\log P$. This is likely because merging counterparts raise $\log P$ during the merging phase. This indicates that it is not necessary to separate the model galaxies by merger experiences after infall when correlating galaxy properties with $\log P$. The weak dependence of $\Phi$ on stellar mass seen in the top right panel of Figure~\ref{merger_infall} results in the small gaps between the two mass groups in $\Phi/\Delta t$. On the other hand, more massive galaxies show larger $\phi/\Delta t$ as already seen in the bottom right panel of Figure~\ref{merger_infall}.

\subsubsection{Dependence of galaxy properties on $\log P$}
The previous section demonstrates a strong correlation between perturbation index and the cumulative angular changes of galaxy spin vectors. It has also been suggested that tidal perturbation accounts for mass loss or morphological transformation in cluster environments by disturbing internal kinematics~\citep[e.g.][]{moore96,moore98,gnedin03b}. Therefore, we investigate the effects of perturbation within the cluster on galaxy properties in this section. Figure~\ref{perturbation2} shows the rate of change for stellar mass, stellar mass density within $R_{\rm eff}$, size of $R_{\rm eff}$, and $V_{\theta}/\sigma$ as a function of $\log P$. The satellite galaxies are grouped into two groups by their final stellar mass, below and above $\log M_{*}/M_{\odot}=10.5$. The hatched regions mark the $16^{\rm th}-84^{\rm th}$ percentiles in the distribution of each parameter and the solid lines denote medians for all the galaxies in each subsample of mass.

Most galaxies are located at the center of their halos, and thus their stellar components will not be significantly stripped unless they are exposed to strong tidal fields for a long time~\citep[e.g.][]{smith16}. In panel (a) of Figure~\ref{perturbation2}, a higher mass stripping rate comes with stronger perturbation in the low mass subsample. On the other hand, there is a great deal of scatter in the high mass subsample. This is not just a stochastic effect due to the scatter increasing with $\log P$. In fact, this is mainly caused by the fact that the stellar mass loss exceeds mass accretion via mergers in many galaxies, finally placing the galaxies in the low mass group at $z=0$. The galaxies in which mergers compensate for mass stripping or even increase the stellar mass end up in the massive group. This behavior results in the weaker correlation between mass change rates and $\log P$ in the more massive group than that of less massive group.  

It is well known that the size of galaxies increases with decreasing redshifts for a given stellar mass~\citep{ferguson04,fan10,dutton11,huertas-company13,morishita14,vanderwel14,shibuya15,anderson16,allen17,furlong17,paulino-afonso17,genel18}, essentially leading to a decline in the stellar mass density of galaxies. The negative offset in panel (b) and positive offset in panel (c) demonstrates the same evolutionary trends. Most satellite galaxies show an increase of $R_{\rm eff}$ after infall and accordingly a decrease of stellar mass density. However, dependence on $\log P$ is different between the specific change rates of density and size. Stellar mass density changes show no dependence on $\log P$ but its scatter becomes larger with increasing and decreasing $\log P$, implying that stochastic effects mainly drive the scatter. On the other hand, larger scatters and smaller increasing rates are seen along increasing $\log P$ in the specific change rates of size. This behavior seems to be consistent with that of the specific change rates of stellar mass, indicating that stellar mass changes mainly govern the dependence of the specific change rate of $R_{\rm eff}$ on $\log P$. Notable features in panels (b) and (c) are the separation of the two mass groups, which is not obvious in other panels. This means that galaxy's stellar mass \st{is} more directly control{\color{red}s} its size or density evolution than environments do.

\begin{figure*}
\centering 
\includegraphics[width=1\textwidth]{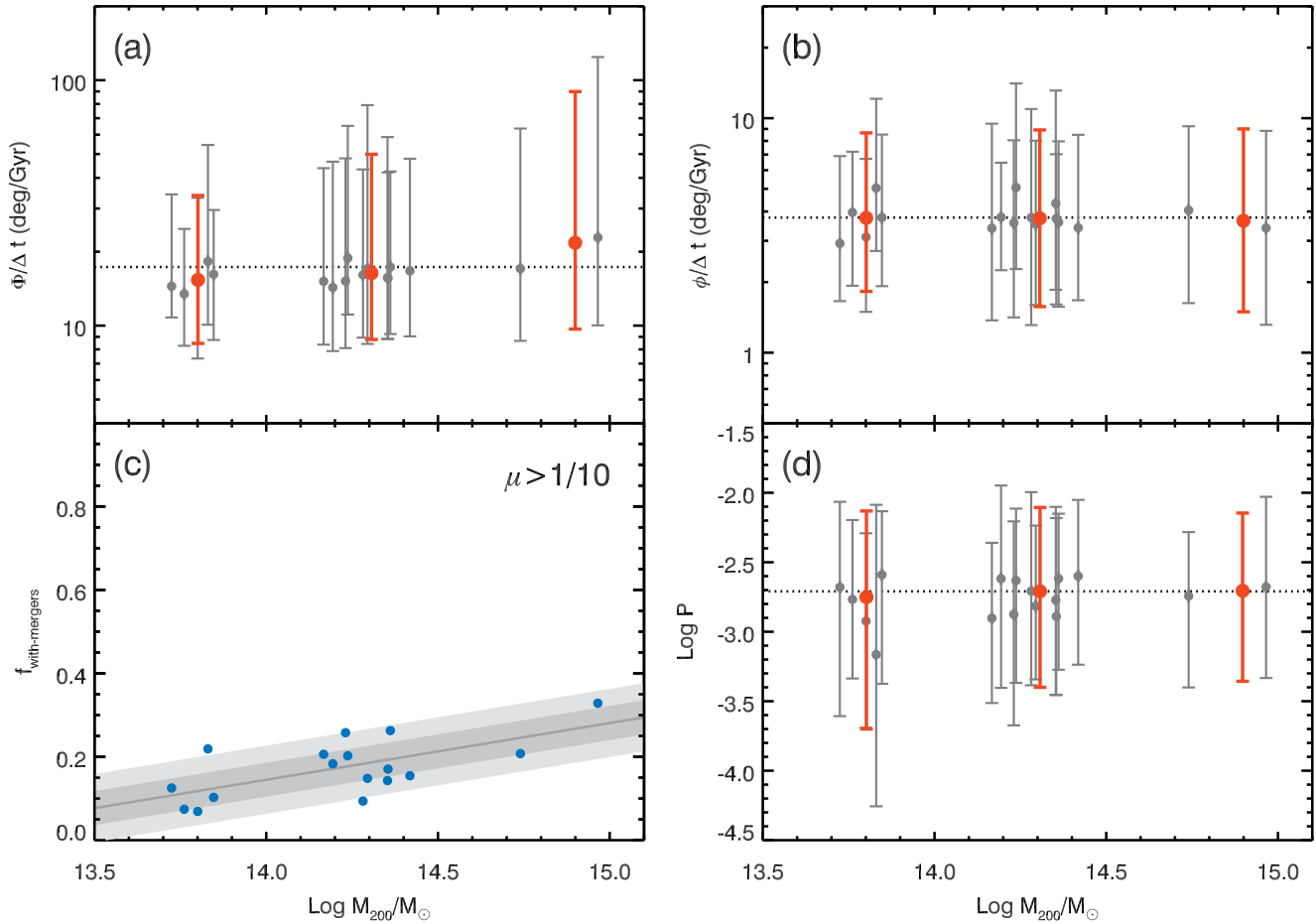}
\caption{(a) The distribution of $\Phi/\Delta t$ of satellite galaxies in the model clusters, where $\Delta t=t_{\rm infall}-t_{z=0}$, as a function of final cluster mass $M_{200}$. The vertical bars with filled circles indicate $16^{\rm th}$, $50^{\rm th}$, and $84^{\rm th}$ percentile distribution. The grey symbols are for individual clusters and red ones are for clusters in three mass ranges $\log M_{200}/M_{\odot}=13.5-14.0$, $14.0-14.5$, and $14.5-15.0$. The horizontal dotted line marks the median $\Phi/t_{\rm infall}$. (b) The distribution of $\phi/\Delta t$ of satellites in clusters. The horizontal dotted line mark the median of $\phi/\Delta t$ of satellite galaxies in all the clusters. (c) The fraction of galaxies that are involved in mergers with $M_2/M_1>1/10$ after infall. The grey solid lines and the shades display the linear fit, $1\sigma$, and $2\sigma$ distribution of the blue filled circles. (d) The distribution of the perturbation index of satellites in clusters. The horizontal dotted line mark the median of the perturbation indices of satellite galaxies in all the clusters}
\label{m_trend}
\end{figure*}

Panel (d) shows the change rate of the specific angular momentum of stellar components $|\vec{h}|$ at $r<R_{\rm eff}$ after infall. The medians of all the cases are smaller than 0, as shown in panel (c) of Figure~\ref{gal_prop}, indicating rotation weakening over time. A weak trend is visible especially among the low-mass galaxies, in qualitative agreement with~\cite{choi18}. Meanwhile, its scatter becomes notably dominant with increasing $\log P$, implying that perturbation is possibly one of parameters driving varying changes of internal kinematics. As shown in the upper panel of Figure~\ref{perturbation_hist}, more galaxies experience mergers which effectively disturb internal kinematics of galaxies after infall with increasing $\log P$, accordingly showing the larger scatters. Summing up, the perturbation index is strongly correlated with the changes of mass and the scatter of the specific angular momentum changes, but does not directly affect size and density \st{changes} of satellite galaxies. Between galaxy mass and environment, the former appears to play more critical role to galaxy size and density, indicating the importance of internal processes for size and density evolution.

\subsection{Survivors vs destroyed satellites}
This study only considered satellite galaxies that survive until $z=0$ in galaxy clusters. However, many galaxies disappear before $z=0$ by merging into other galaxies or being tidally disrupted after infall. The initial mass ratio of subhaloes to host haloes and initial orbital circularity are thought to be two main parameters governing merging timescales~\citep[e.g.][]{boylan-kolchin08}. We compared the values of these two parameters for the two separate populations, i.e. the merged and surviving populations. We find they are similar in initial mass ratio, but the merged one has an orbital circularity $\eta$ (on first passage) that is lower than that of the surviving population (median $\eta=0.47$ for the surviving population while median $\eta=0.41$ for the merged one). Therefore, the merged satellites more closely approach the inner cluster and may experience stronger dynamical friction and tidal perturbations. Consequently, the merged satellites also undergo larger spin orientation changes than the surviving satellites do ($\sim7$ deg/Gyr of median $\phi/\Delta t$ and $\sim19$ deg/Gyr of median $\Phi/\Delta t$ in the merged population versus $\sim4$ deg/Gyr of median $\phi/\Delta t$ and $\sim17$ deg/Gyr of median $\Phi/\Delta t$ in the surviving one). Like in panel (b) of Figure~\ref{gal_prop}, orbit is once again highlighted as a primary parameter governing spin orientation changes of satellite galaxies in clusters.

\subsection{Cluster-to-cluster variation}
So far, we have studied the angular change in satellite spin vectors in the 17 model clusters combined together, without examining possible cluster-to-cluster variations. As described in \S2.1, however, the model clusters used in this study have varied virial masses, $\log M_{200}/M_{\odot}\sim13.7-15.0$ at $z=0$, and could also vary in their environment, merger-history, and formation time~\citep[see][and references therein]{lee17}. In this section, we briefly examine cluster-to-cluster variations. 

Figure~\ref{m_trend} displays, for each individual cluster, (a) the time averaged cumulative angular changes of spin vectors integrated for $\Delta t=t_{\rm infall}-t_{z=0}$, (b) the time averaged net angular changes of spin vectors after infall over $\Delta t$, (c) the fractions of galaxies that are involved in mergers of mass ratio above 1/10 after infall, and (d) the perturbation index after infall as a function of the virial mass of the clusters at $z=0$. The grey filled circles denote the medians from the distribution of satellite galaxies in each cluster, and the associated error bar indicate the $16^{\rm th}-84^{\rm th}$ percentile distribution. The red vertical bars show the distributions of the measurements of the satellite galaxies in the clusters in bins of $\log M_{200}/M_{\odot}=13.5-14.0$, $14.0-14.5$, and $14.5-15.0$ at $z=0$. 

There is a slight hint of increasing $\Phi/\Delta t$ with increasing cluster mass in panel (a), but this does not seem to be statistically significant. In panel (b), $\phi/\Delta t$ does not show a dependence on cluster mass at all. On the other hand, the fractions of the galaxies with mergers of $\mu>1/10$ show a relatively clear trend with increasing cluster mass in panel (c). Satellites in more massive clusters are more likely to merge with each other after infall. This result seemingly contradicts the common sense point-of-view that more massive clusters are not favored environments for satellite-satellite mergers due to higher peculiar velocities. This is however not necessarily the case in evolving clusters~\citep[see e.g.][]{gnedin03b,yi13}. Theoretical studies have demonstrated that more massive halos are assembled later~\citep[e.g.][and references therein]{neistein06}. Therefore, more massive clusters are more likely to be less relaxed, and harbor subgroups that fell into them recently. Satellites in subgroups have enhanced chances over cluster satellites to merge with their old neighbors before the subgroups are broken up within the cluster. This higher fraction of merger events seems to be one of causes inducing the slightly larger $\Phi/\Delta t$ in more massive haloes in panel (a). In panel (d), satellite galaxies experience almost the same degree of perturbation after infall into the clusters of varying virial mass. This is mainly because all the haloes essentially have the same mean density due to the definition of $M_{200}$ and $R_{200}$. Summing up, final cluster mass does not have a significant impact on the resulting angular changes of satellite spin vectors.

\section{Spin vector changes in controlled simulation}

Cosmological simulations are inherently complex, and an individual galaxy may suffer multiple mechanisms, often in parallel, that can change the spin vector. Therefore we conduct a series of controlled simulations to complement our cosmological simulations. By using controlled simulations, we are able to systematically vary one parameter, while fixing the other parameters. In this way, we may gain a deepened understanding of the dependency of spin vector changes to an individual parameter.

The controlled simulations are conducted with the Treecode algorithm \gf\ which operates primarily using the techniques described in ~\citet{Hernquist89}. The Treecode allows for rapid calculation of gravitational accelerations. We have previously applied this code in multiple studies of the impact of the cluster environment on galaxies~\citep{smith10,smith12a,smith12b,smith13,smith15}, in which a more detailed description of the code can be found. In controlled simulations all of our computing power can be focussed on a single model galaxy. Thus we can reach significantly higher spatial and mass resolution than is possible in the cosmological simulations. For example, all our model galaxies in the controlled simulations have a gravitational softening length of 25pc, and a particle mass of $3.3\times10^4M_{\odot}$ and $10^4M_{\odot}$ for dark matter and stars respectively, which are better compared with up to $\sim1$kpc of gas cell resolution, and a particle mass of $8\times10^7M_{\odot}$ and $5\times10^5M_{\odot}$ for dark matter and stars in the cosmological zoomed simulations. This allows us to test if our conclusions, derived from the lower resolution cosmological simulations, are robust to increases in resolution. 

\begin{figure*}
\centering 
\includegraphics[width=1\textwidth]{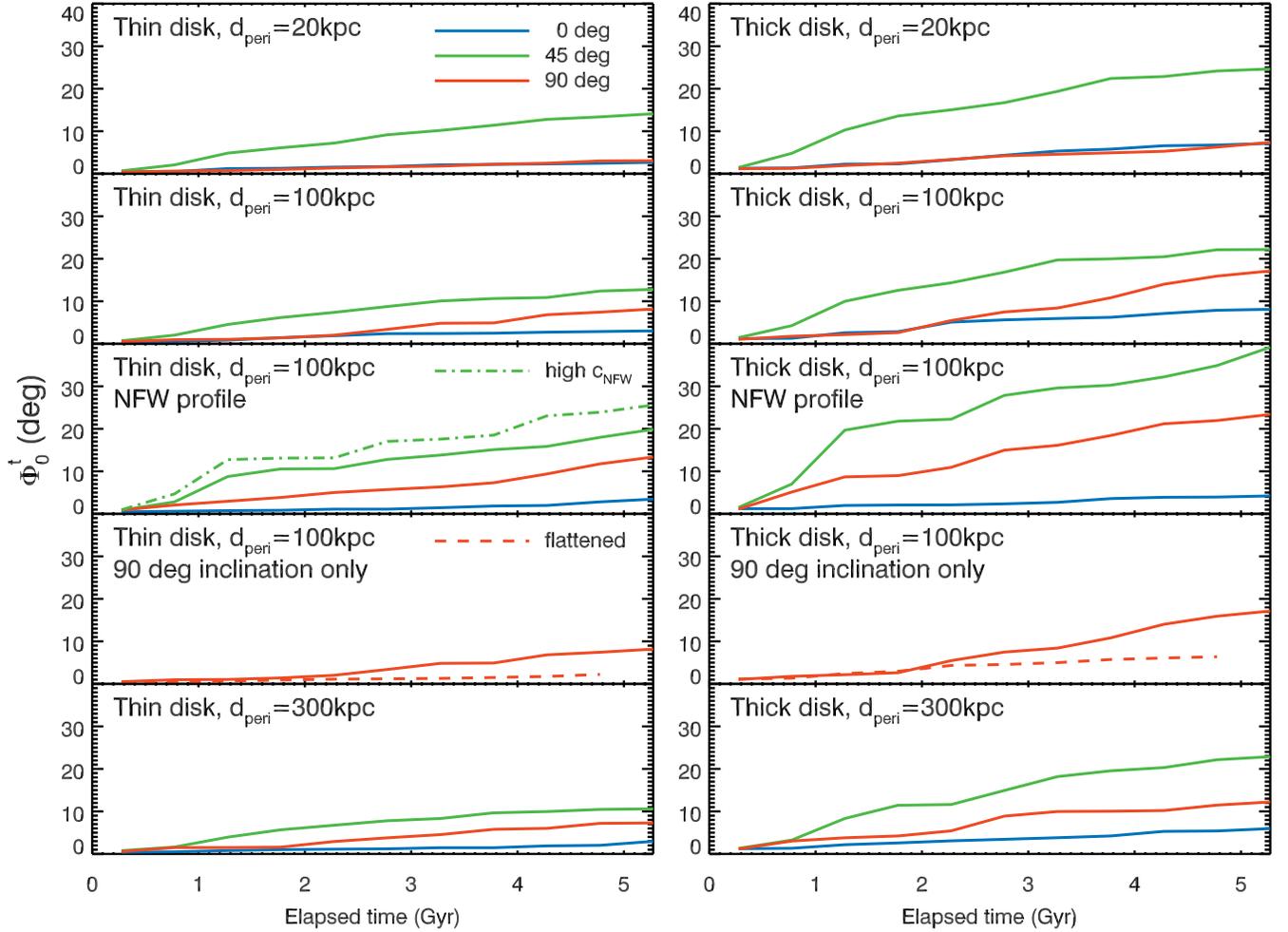}
\caption{Cumulative angular changes of galaxy spin orientation integrated from $t=0$ to given elapsed time in a set of ideal simulations. The left and right panels are for thin and thick disk galaxies, respectively. The line colors indicate the initial inclination between disks and orbital planes. The panels in the third row present the $\Phi_0^t$ of galaxies in halos following the NFW profile while the other rows are for halos following the Plummer profile. The green dot-dashed line in the left panel of the third row is for the galaxy initially having 45 deg of inclination in halos with higher $c_{\rm NFW}$ than the other NFW cases. The fourth row panels display the impact of cluster halo shapes on the galaxies with a 90 deg initial inclination. In the panels, the red solid and dashed lines are for the spherical and flattened potentials, respectively (see text for details).}
\label{ideal}
\end{figure*}

\subsection{Setup}

With the controlled simulations, we focus on studying how the main cluster's gravitational potential can drive changes in the spin vector of a galaxy that orbits within it. Although we vary various parameters controlling the cluster potential's properties (see below), we always ensure that the cluster radial density profile contains 2$\times$10$^{14}$~M$_\odot$ within 1~Mpc. In general a spherical cluster potential is considered, unless otherwise stated. In each simulation, we initially place a model galaxy at a clustocentric radius of 650~kpc, and we vary the initial tangential velocity in order to produce the required pericentric radius. Galaxy orbits are conducted for 5.5~Gyr. In this time, each model galaxy conducts three consecutive pericentre and apocentre passages. The cluster is modelled with a static potential in our controlled simulations. Thus the lack of dynamical friction between the galaxy and cluster could lead to an underestimate of tidal stripping, but our approach has the advantage of allowing us to precisely control the orbits of the model galaxies.

We chose to model a fairly massive dwarf galaxy as such a galaxy would be present, but poorly resolved, within our cosmological models. Each model galaxy has an NFW dark matter halo~\citep{navarro97} consisting of 3$\times$10$^6$ dark matter particles, with a total Virial mass of 1$\times$10$^{11}$~M$_\odot$, a halo concentration of 15, and a Virial radius of 95~kpc. There is also a stellar disk consisting of 1$\times$10$^5$ star particles, with an exponential profiles, and an effective radius of 1.7~kpc. We choose to model an early type disk dwarf galaxy, and so do not include a gas component into our models, in order to focus our study on the impact of cluster tides on the pre-existing stellar disk of a cluster galaxy. Model galaxies are initially evolved for 0.5~Gyr in isolation, to ensure they are well relaxed before introducing them into the cluster potential.

\subsection{Result}

We conduct a limited parameter study of 29 models, varying one parameter at a time, while fixing the others. In the following, we describe each parameter that was varied, and describe that parameters impact on the angular changes of the satellite spin orientation. The resulting changes in spin orientation is shown Figure~\ref{ideal}. The measurement of the change in the spin parameter is calculated in the same way as was done for the cosmological simulations. As a result, we do not see sudden changes in spin vector when a galaxy passes pericentre, as the measured spin parameter value is smoothed over half-gigayear steps. We note that we do not see tidal stripping of the stellar disk in any of our simulations. Therefore measured changes in the spin vector occur purely as a result of modification of the spin vectors of star particles within the stellar disk, and not as a result of the removal of star particles with particular spin vector properties. We also measured the contribution of non-perturbation induced changes to the spin orientation by evolving the galaxy model in complete isolation for 5.5 Gyr. We note that its cumulative spin orientation change is no more than ~3 deg at the final time step of the simulation.

\begin{itemize}
\item {\it{Disk Thickness:}} We vary the vertical velocity dispersion of each model's stellar disk to produce a thin disk model and a thick disk model, with an axial ratio of 0.05 and 0.2 when seen edge-on. Thicker disks correspond to those with lower $V_{\theta}/\sigma$. In Figure~\ref{ideal}, the effects of varying stellar disk thickness on spin angle change can be seen by comparing between the left column for thin disk and the right column for thick disk of a row. In general, we see that the thick disk model suffers roughly 1.5 to 2 times larger changes in spin vector than the thin disk model. Thus the disk thickness is an important parameter governing changes in spin vector, and this conclusion is entirely consistent with the conclusions drawn from our cosmological models.

\item {\it{Disk Inclination:}} We vary the inclination of the stellar disk with respect to its orbital plane and consider three values; 0 degrees (the disk lies in the orbital plane), 45 degrees, and 90 degrees (the disk plane is perpendicular to the orbital plane). The effect of disk inclination can be seen by comparing the red, blue and green lines in a panel (see legend). Disk inclination is also a very important parameter governing changes in spin vector. In general, the 0 degree inclination models have the smallest changes in spin vector, the 90 degree inclination models have intermediate strength changes, and the 45 degree inclination models have the largest changes. The 45 degree inclination models often suffer more than three times the spin vector change of the 0 degree inclination models. Logically, the angle of the disk to the cluster potential's isocontours ($\theta_{twist}$) will be an important factor that gives rise to torquing on the disk, especially near pericentre where the tides are most strong. In the 0 degree inclination case, $\theta_{twist}$ will always be zero, and so there is no torquing. However, in the the 90 degree inclination model, the torquing varies in strength and direction along the orbit. In the 45 degrees inclination case, $\theta_{twist}$ is always at least 45 degrees resulting in stronger torquing. Additionally, orbital precession must play a role, as this causes the direction of the torquing to change with each orbital period. For the orbits considered here, smaller pericentres orbits result in less precession. As a result, the 0 and 90 degree lines (blue and red respectively) are close together for small pericentre orbits and further away for larger pericentre orbits. We also conducted counter-rotating models with a 180 degree inclination. These models show similar but marginally increased changes in spin orientation ($<$30$\%$ increase) than those of the 0 degree inclination models. However, this is not significant compared with the differences between the 0 degrees, 45 degrees, and 90 degrees models. We expect that, in the cosmological models, the disk inclination may be quite randomly orientated with respect to their orbital planes. Thus the disk inclination dependency will mostly become a source of scatter in the dependency found on other parameters in the cosmological models.

\item {\it{Pericentre Distance:}} We vary the initial tangential velocity in order to control the pericentre distance of the orbit. We consider three values of pericentre distance; 20 kpc, 100 kpc, and 300 kpc corresponding to R$_{\rm{peri}}$/R$_{\rm{vir}}$=0.024, 0.118, 0.353 for our standard density profile (see below for details). Although 20 kpc appears quite extreme, because we use a standard density profile that is cored, we find we must vary the pericentre distance considerably to see any resulting differences in the spin vector. This can be seen in Figure~\ref{ideal} by comparing between the top two rows and the bottom row, between which only the pericentre distance $d_{\rm peri}$ is varied. Closer pericentre passages result in increasing changes in the spin vector, in consistent with panel (b) of Figure~\ref{gal_prop}, but the effect is weaker than the dependency seen on the previously described parameters. However, we note that this weak dependency may also be an artifact of our choice of a cored density profile for our standard cluster model, which we will remedy below.

\item {\it{Flattening of the Cluster Potential:}} For the previously described parameters, we chose a Miyamoto-Nagai potential~\citep{miyamoto75} for the cluster gravitational potential. This type of potential is controlled by three parameters; a mass M$_{MN}$, and two scalelengths $a$ and $b$. One advantage of using this potential is the ease by which this density distribution can be flattened. We use M$_{MN}$=2.78$\times$10$^{14}$~M$_\odot$, and simply by setting $a=0$ and $b$=500~kpc, we produce a spherical potential\footnote{A spherical Miyamoto-Nagai potential is identical to a Plummer model with Plummer scalelength=$b$}. This is the spherical cluster potential used for the controlled simulation tests described above. However, we also test the effects of a non-spherical cluster halo by setting $a$=$b$=500~kpc. This results in a quite flattened density distribution, with an axial ratio of 0.55 when seen edge-on. We increase M$_{MN}$ by a factor of roughly two, in order to ensure that there is 2$\times$10$^{14}$~M$_\odot$ within 1~Mpc (as is the case for all our spherical cluster models). We choose a 90 degree inclination which, due to our set-up, means the orbit of the galaxy is perpendicular to the plane about which the cluster potential is flattened. It is also the only inclination we found that result in each pericentre being similar (orbital parameters may vary significantly with time in a flattened cluster potential). We evolve each model until it completes 3 pericentre passages, like in the spherical potential cases, although here the orbital period is slightly shorter. The results are shown in row 4 of Figure~\ref{ideal}. The change in spin vector is even weaker for the flattened cluster potential model (red dashed line) than in the spherical potential (solid line). This is likely a result of our choice of orbit. We chose this orbit as it resulted in a quite stable orbit where each pericentre distance was similar. But this particular orbit shows almost no precession, meaning the disk always has a similar angle to the cluster potential's isocontours at pericentre, and this results in reduced torquing of the disk. In summary, it is challenging to isolate the impact of cluster potential flattening on spin vector change as it has multiple knock-on effects on a galaxy?s orbit. However, we can conclude that we do not see clear evidence for a strong enhancement in spin vector changes as a result of cluster potential flattening.

\item {\it{Cluster Radial Density Profile:}} As noted above, we chose our standard cluster potential to be a Miyamoto-Nagai potential. This potential has a cored density profile which is not a good match to the cuspy NFW density profiles found in cosmological simulations. Therefore, we also conduct an additional set of simulations using an NFW cluster potential. Our standard NFW model has a Virial mass of 2.5$\times$10$^{14}$~M$_\odot$, a concentration of 4, and a Virial radius of 1290~kpc. However, we also consider an NFW cluster model with a higher concentration of 8. The results of this simulation set can be seen in row 3 of Figure~\ref{ideal}. Comparing the solid lines with those in row 2, it is clear that the NFW halo has caused an increase in change in the spin vector, compared to the Miyamoto-Nagai profile cluster model. The effect is generally small but non-negligible. The dash-dotted line in the left panel of row 3 indicates that the increased halo concentration also results in a small additional increase in the spin vector change. This is likely because by increasing the mass near the cluster centre, the gravitational tides near pericentre are stronger, resulting in stronger spin vector chane. For the cosmological simulations, these results suggest that cluster to cluster variations in halo density profile and concentration could be additional parameters influencing disk spin changes. These parameters could result in additional scatter in the dependencies that have been found in the cosmological simulations.
\end{itemize}

\section{Summary and Conclusion}
This study looked into the angular stability of spin vectors in dense environments. In relatively less dense environments, theoretical studies have suggested that galaxy spin vectors are initially aligned with the direction of nearby filaments, but are able to be swung by hierarchical processes~\citep[e.g.][]{codis12,dubois14,codis15b}. This theoretical prediction has been confirmed by a series of empirical studies which have shown that late types have spin vectors more likely to be aligned with nearby filaments than those of early types galaxies which probably underwent several mergers in the past~\citep[e.g.][]{tempel13b,zhang13,zhang15,hirv17}. 

As the densest structures in the Universe, galaxy clusters are located at the nodes of filamentary structures which arrive from various directions. The degree of spin alignment of cluster galaxies is thus found to vary between clusters, with respect to their surrounding large scale structures~\citep{aryal04,aryal05,hwang07,kim18}. Meanwhile, galaxy clusters are dynamically harsh environments where strong tidal forces can result in mass loss, distortion of disks, or even morphology transformation of satellite galaxies~\citep[e.g.][]{moore96,moore98,gnedin03a,mastropietro05,aguerri09,smith10,smith12a,smith15,bialas15}. Furthermore, mergers can take place between satellite galaxies in subgroups if clusters are not well relaxed. Therefore, a question arose in a general sense as to how much the orientation of galaxy spin vectors is changed in dense environments. 

To answer the question, we utilize a set of zoomed cosmological hydrodynamical simulations of clusters \yzics~\citep{choi17}, supplemented with a set of idealized simulations devised to study the impact of various parameters on spin vector changes. We parameterized the angular changes of spin vectors after infall into cluster environments in two ways: their net changes between two epochs $\phi$, and their cumulative changes over a period of time $\Phi$. The degree of perturbation induced by tidal forces from neighboring galaxies is also quantified in terms of the Tidal perturbation index $\log P$. We then examine the correlations of various physical parameters with the resulting angular changes in galaxy spin vectors in dense environments. Based on the cosmological simulations, our conclusions are as follows: 

\begin{itemize}
\item  Cluster satellites undergo large changes in the direction of their spin vectors when their orbits take them to small clustocentric distances,  and when their disks are more dispersion supported at the time of their infall into the cluster. Changes in the direction of galaxy spin vectors are closely connected with changes in the amplitude of the spin. More massive galaxies have larger $\Phi$, but this is primarily because they tend to have more dispersion supported kinematics.

\item Galaxy mergers are also an important factor in enhancing changes in spin vector direction, resulting in $\Phi$ and $\phi$ values that are roughly a factor of two greater than the no merger case at fixed infall epochs, clustocentric distances at $z=0$, and final stellar mass. 

\item Because close passages past the cluster centre can result in changes, we find a clear trend for $\Phi$ to increasing with decreasing 3D clustocentric distance at $z=0$. However, $\phi$ is a better tracer of what we would observe at any instant, and this shows a much weaker correlation with 3D clustocentric distance. And this trend would be even weaker if we were to use projected distances to the cluster as is observed. Most of satellite galaxies have $\phi$ less than 30 deg after infall at any location in clusters at $z=0$, which indicates that pre-infall orientations are not strongly erased by the cluster environment. Furthermore, $\phi$ is almost always less than $\Phi$, which shows that frequent changes in the direction of a galaxy's spin vector are later canceled out (i.e. dominated by motion involving wobbling back and forth, rather than tending to change in any one direction). The median of time averaged cumulative angular change $\Phi/\Delta t$ is $\sim17$ deg/Gyr, four times larger than time averaged net angular changes $\phi/\Delta t$ after infall. The input of non-physically induced changes in the cumulative angular changes is estimated to be $\sim$1/3 of the $\Phi/\Delta t$.

\item The tidal perturbation index $\log P$ shows a clear correlation with $\Phi/\Delta t$, but a much weaker correlation with $\phi/\Delta t$. This is because the direction of the net tidal force changes as satellites orbiting in clusters, which is likely also an important factor for wobbling motion of galaxy spin vectors. When the tidal perturbation index is high, galaxies lose stellar mass. 


\item We search for cluster-to-cluster variations between our model clusters, but do not see a strong indication of differences, except a tendency for more massive clusters to contain a higher frequency of mergers with mass ratio greater than 1:10. However, this is not sufficient to result in a significant enhancement in the changes in spin vector direction. 
\end{itemize}

Cosmological zoomed simulations describe the evolution of galaxies in evolving environments in a realistic manner. However a consequence of this realism is that often multiple phenomena occur, sometimes in parallel. Therefore, we also carried out a set of controlled simulations to supplement the cosmological zoomed simulations. In these simulations, we attempt to pick apart how the cluster potential can transform disk spin direction, as a function of disk thickness, disk inclination, pericenter distances, the shape of cluster potential, and the density profile of cluster potential. These controlled simulations show that thicker disks, which correspond to lower $V_{\theta}/\sigma$, and shorter pericenter distances lead to larger spin vector changes in cluster environments. Spin vectors are most stable when the disk is inclined such that it lies within the galaxy's orbital plane. When the disk inclination is 45 degrees to the orbital plane, the largest changes in spin vector direction occur. Steeper gradients of the cluster potential (i.e. larger concentration index of the NFW profile) induce larger changes in galaxy spin vector direction. {\st The non-sphericity of the cluster potential was also found to be a factor, with the more flattened cluster resulting in smaller angular changes.
Cosmological zoomed simulations describe the evolution of galaxies in evolving environments in a realistic manner. However a consequence of this realism is that often multiple phenomena occur, sometimes in parallel. Therefore, we also carried out a set of controlled simulations to supplement the cosmological zoomed simulations. In these simulations, we attempt to pick apart how the cluster potential can transform disk spin direction, as a function of disk thickness, disk inclination, pericenter distances, the shape of cluster potential, and the density profile of cluster potential. These controlled simulations show that thicker disks, which correspond to lower $V_{\theta}/\sigma$, and shorter pericenter distances lead to larger spin vector changes in cluster environments. Spin vectors are most stable when the disk is inclined such that it lies within the galaxy's orbital plane. When the disk inclination is 45 degrees to the orbital plane, the largest changes in spin vector direction occur. Steeper gradients of the cluster potential (i.e. larger concentration index of the NFW profile) induce larger changes in galaxy spin vector direction. Cosmological zoomed simulations describe the evolution of galaxies in evolving environments in a realistic manner. However a consequence of this realism is that often multiple phenomena occur, sometimes in parallel. Therefore, we also carried out a set of controlled simulations to supplement the cosmological zoomed simulations. In these simulations, we attempt to pick apart how the cluster potential can transform disk spin direction, as a function of disk thickness, disk inclination, pericenter distances, the shape of cluster potential, and the density profile of cluster potential. These controlled simulations show that thicker disks, which correspond to lower $V_{\theta}/\sigma$, and shorter pericenter distances lead to larger spin vector changes in cluster environments. Spin vectors are most stable when the disk is inclined such that it lies within the galaxy's orbital plane. When the disk inclination is 45 degrees to the orbital plane, the largest changes in spin vector direction occur. Steeper gradients of the cluster potential (i.e. larger concentration index of the NFW profile) induce larger changes in galaxy spin vector direction. The non-sphericity of the cluster potential was also found to be a factor, with the more flattened cluster resulting in smaller angular changes.

An important implication of this study is that clusters do not effectively erase the signatures of pre-infall spin orientation of satellite galaxies. Mergers reorient spin vectors, but mergers are not frequent between satellites. Tidal perturbation swings spin vectors no less than mergers, but its net effect is not significant. 
These results give us theoretical justification to correlate the spin orientation of galaxies and the direction of nearby large scale structures in dense environments. Besides, these results reconcile~\citet{aryal04,aryal05} and~\citet{kim18} who found varying spin alignment nature between cluster satellites and nearby large scale structures by suggesting that the orientation of galaxy spin vectors is predominantly determined before cluster infall, and well preserved for many gigayears after cluster infall. As such, the spin vector orientation of cluster satellite galaxies encodes valuable information on large scale structure formation and evolution.
 
\section*{acknowledgments}
We thank the anonymous referee for constructive comments that improved the clarity of the manuscript significantly. In particular, Section 3.5 is a result of the communication with the referee.
H.J. acknowledges support from the Basic Science Research Program through the National Research Foundation (NRF) of Korea, funded by the Ministry of Education (NRF-2013R1A6A3A04064993). SKY acknowledge the support from the Korean National Research Foundation of Korea (NRF-2017R1A2A1A05001116). This study was performed under the umbrella of the joint collaboration between Yonsei University Observatory and the Korean Astronomy and Space Science Institute. The supercomputing time for numerical simulation was kindly provided by KISTI (KSC-2014-G2-003), and large data transfer was supported by KREONET, which is managed and operated by KISTI.

\end{document}